# ON THE FLOW OF TIME[1]

George F R Ellis,
Mathematics Department, University of Cape Town
Rondebosch 7701, Cape Town, South Africa.

**Abstract:** *Current theoretical physics suggests the flow of time is an illusion: the entire universe just is, with no special meaning attached to the present time. This paper points out that this view, in essence represented by usual space-time diagrams, is based on time-reversible microphysical laws, which fail to capture essential features of the time-irreversible nature of decoherence and the quantum measurement process, as well as macro-physical behaviour and the development of emergent complex systems, including life, which exist in the real universe. When these are taken into account, the unchanging block universe view of spacetime is best replaced by an evolving block universe which extends as time evolves, with the potential of the future continually becoming the certainty of the past; spacetime itself evolves, as do the entities within it. However this time evolution is not related to any preferred surfaces in spacetime; rather it is associated with the evolution of proper time along families of world lines. The default state of fundamental physics should not be taken to be a time irreversible evolution of physical states: it is an ongoing irreversible development of time itself.*

## 1: The flow of time

The most important property of time is that it unfolds. The present is different from both the past and future, which in turn are completely different from each other, the past being fixed and the future changeable. The present is the instant of transition between these two states. The time that is the present at this instant will be in the past at the next instant. This process of coming into being rolls on in an immutable way: while we can influence what happens in time, we cannot influence the way that time itself progresses on. As stated by Omar Khayam,[2] "The Moving Finger writes: and, having writ, Moves on: nor all thy Piety nor Wit, Shall lure it back to cancel half a Line, Nor all thy Tears wash out a Word of it."

It is this fundamental a feature of time that is not captured by today's theoretical physics, even though it is a most obvious and crucial feature of time in everyday life. This disjunction is captured most clearly in Julian Barbour's book *The End of Time* [1], proposing that the flow of time is an illusion. This conclusion is reached because solutions of the Wheeler-de Witt equation, the fundamental equation in quantum gravity for the wave function of the universe (containing all information about the geometry and matter content of the universe), are necessarily time independent. Barbour takes this feature seriously and so arrives at his conclusion.

Similar conclusions arise from more general analyses of the equations of theoretical physics, leading to the (implicit or explicit) "block universe" picture of spacetime, where the present time has no particular meaning: indeed it is often not even represented in the relevant spacetime diagrams. A closely related feature is the crucial question of time irreversibility: the laws of physics, chemistry, and biology are irreversible at the macro scale, as evidenced *inter alia* by the Second Law of Thermodynamics, even though the laws of fundamental physics (the Dirac equation, Schroedinger's equation, Maxwell's equations, Einstein's field equations of gravity, Feynman diagrams) are time reversible. This irreversibility is a key aspect of the flow of time: if things were reversible at the macro scale, there would be no genuine difference between the past and the future, and the physical evolution could go either way with no change of outcome; both developments would be equally determined by the present. The apparent passage of time would have no real consequence, and things would be equally predictable to the past and the future.

So is the feature of the flow of time and the arrow of time a property of the emergence of complexity, true at larger scales but not at micro scales? I will argue no: the time irreversibility and the flow of time are crucially already built into quantum mechanics through the quantum measurement process, however one interprets it: as collapse of the wave function [2], through a many universes approach ([3]:156-159), or through decoherence [4] and/or multiple histories [5]. This then affects all the rest of physics, and hence chemistry and biology, even though there are some circumstances where boundary conditions and local equations of state lead to almost reversible behaviour. The flow of time affects not only physical processes within spacetime, but the very structure of spacetime itself; hence an *Evolving Block Universe* ("EBU") [6] is the appropriate spacetime picture to use, where the present is different from the past and future, and the very nature of the future (including spacetime) is undetermined until the instant it happens.

---

[1] Essay for the Fqxi essay contest on THE NATURE OF TIME.
[2] E. Fitgerald, *The Rubaiyat of Omar Khayyam*. (Penguin 1989), Stanza lxxi.

This has various consequences: it means that attempts to explain the arrow of time or other features of local physics through top-down processes from the future, or a comparison of boundary conditions in the infinite future and past, cannot succeed, for the future does not yet exist and so cannot be taken into account in such calculations. It affects discussions of closed timelike lines and the proposed cosmic censor who may prevent causal violations. It affects attempts to promote the supposed insignificance of the present in theoretical physics to real world situations such as discussions of free will. And it locates the key issue of the flow of time where it belongs: as a local process in the foundations of quantum measurement theory and presumably quantum gravity. Indeed, perhaps taking the flow of time properly into account may be an important step in developing a satisfactory theory of quantum gravity.

**2: The quantum flow of time**

Is the flow of time a purely macro phenomenon, arising through a process of coarse-graining [2] out of a microphysical domain where there is no flow of time and no arrow of time? No, because the quantum measurement process is not time reversible. In the measurement process, the wave function evolves almost instantaneously to an eigenstate (the next measurement will give the same result with complete certainty), but the specific eigenstate is not predictable: the outcome it is only known as it occurs. It is inherently unpredictable, but the probabilities of different outcomes of the evolution during measurement are known, indeed this is the principle outcome of quantum theory. However the unpredictability of quantum theory is far more serious to the past than the future: even the probabilities are unknown. Hence the fundamental irreversibility is already there at the quantum level.

More formally: if a measurement of an observable $A$ takes place at time $t = t_*$, initially the wave function $\psi(x)$ is a linear combination of eigenfunctions $u_n(x)$ of the operator $\tilde{A}$ that represents **A**: for $t < t_*$, the wave function is

$$\psi_1(x) = \Sigma_n \psi_n u_n(x). \tag{1}$$

([3]:5-7). But immediately after the measurement has taken place, the wave function is an eigen-function of $\tilde{A}$: it is

$$\psi_2(x) = a_N u_N(x) \tag{2}$$

for some specific value N. The data for $t < t_*$ do not determine the index N; they just determine that the probability a measurement of A will yield a particular eigenvalue $a_N$ is

$$Prob\ (A=a_N;\ \psi) = |\psi_N|^2. \tag{3}$$

One can think of this as due to the probabilistic time-irreversible collapse of the wave function ([2]:260-263). Invoking a many-worlds description ([3]:156-159) will not help: in the actually experienced universe in which we make the measurement, N is unpredictable. Similarly claiming it is actually due to decoherence [4] also makes no practical difference: we still can't uniquely predict the future from the past, the best we can do is the prediction given in eqn. (3). Thus the initial state (1) does not uniquely determine the final state (2); and this is not due to lack of data, it is due to the foundational nature of quantum interactions. You can predict the statistics of what is likely to happen but not the unique actual physical outcome, which unfolds in an unpredictable way as time progresses; *you can only find out what the outcome is after it has happened*. Furthermore, in general *the time $t_*$ when an event occurs is also not predictable from the initial data*: you don't know when `collapse of the wave function' (the transition from (1) to (2)) will happen (you can't predict when a specific excited atom will emit a photon, or a radioactive particle will decay).

*We also can't retrodict to the past at the quantum level*, because once the wave function has collapsed to an eigenstate we can't tell from its final state what it was before the measurement. You cannot retrodict uniquely from the state (2) immediately after the measurement takes place, or from any later state that it then evolves to via the Schrodinger equation at later times t > t$_*$, because knowledge of these later states does not suffice to determine the initial state (1) at times t < t$_*$: the set of quantities $\psi_n$ are not determined by the single number a$_N$. There is no analogue of (3) that even probabilistically determines $\psi_1$ from u$_N$, nor any equation determining t$_*$ retroactively. The situation is completely time asymmetric. Similarly, decoherence is a process that takes place in one direction in time; it is not time reversible.

**Key feature: Quantum physics introduces process, unpredictability, and an arrow of time: the flow of time is built into its deep nature.**

One can actually see the process of collapse taking place through high precision experiments [7]. Whatever underlying theory one may have for what happens, the fact that it effectively takes place is now a solid experimental result.

**3. A Realistic Space-Time Picture**

The time reversible picture of fundamental physics underlying the block universe viewpoint simply does not take these kinds of phenomena into account, specifically because it does not take cognisance either of the progress of time in quantum measurements, or of how complex phenomena arise from the underlying micro-physics, with the emergence of the macroscopic arrow of time as a major feature of chemistry, biology, and human life. It does not take seriously the physics and biology of the real world but rather represents an idealised view of things which is reasonably accurate in certain very restricted situations (for example laboratory experiments where all external influences except that of the experimenter are excluded).

In order to take into account the flow of time that actually occurs both in the quantum world and at macro scales, we need to modify the block universe pictures so as to adequately represent causation in these contexts. How can we envisage spacetime and the objects in it as time unrolls? A way to do this is to consider an *Evolving Block Universe* model of reality, with spacetime ever growing and incorporating more events as time evolves along each world line [6].

To motivate this, consider the following scenario: A massive object has rocket engines attached at each end to make it move either left or right. The engines are controlled by a computer that decides what firing intervals are utilised alternately by each engine, on the basis of a non-linear time dependent transformation of signals received from a detector measuring particle arrivals due to random decays of a radioactive element. These signals at each instant determine what actually happens from the set of all possible outcomes, thus determining the actual spacetime path of the object from the set of all possible paths (Figure 1). This outcome is not determined by initial data at any previous time, because of quantum uncertainty in the radioactive decays. As the objects are massive and hence cause spacetime curvature, the spacetime structure itself is undetermined until the object's motion is determined in this way. Instant by instant, the spacetime structure changes from indeterminate (i.e. not yet determined out of all the possible options) to definite (i.e. determined by the specific physical processes outlined above). Thus a definite spacetime structure comes into being as time evolves. It is unknown and unpredictable before it is determined.

Something essentially equivalent has already occurred in the history of the universe. According to the standard inflationary model of the very early universe, we cannot predict the specific large-scale structure existing in the universe today from data at the start of the inflationary expansion epoch, because density inhomogeneities at later times have grown out of random quantum fluctuations in the effective scalar field that is dominant at very early times:

> "Inflation offers an explanation for the clumpiness of matter in the universe: quantum fluctuations in the mysterious substance that powered the [inflationary] expansion would have been inflated to astrophysical scales and therefore served as the seeds of stars and galaxies" ([8]; see [9] for details).

It follows that *the existence of our specific Galaxy, let alone the planet Earth, was not uniquely determined by initial data in the very early universe*. The quantum fluctuations that are amplified to galactic scale are unpredictable in principle. Thus *spacetime evolution is not predictable even in principle* in physically realisable cases. The outcome is only determined as it happens.

The Evolving Block Universe model of spacetime represents this kind of situation, showing how time progresses, events happen, and history is shaped. Things could have been different, but second by second, one specific evolutionary history out of all the possibilities is chosen, takes place, and gets cast in stone. The way this happens is illustrated in the space time diagram given in Figure 2. This idea was proposed many years ago by Broad [10], but has not caught on in the physics community. The proposal here is that it represents the foundational nature of physics as well as of biology. The more usual block spacetime diagrams simply omit an essential feature of physical reality.

**3.1 General Relativity**
In Figure 2, the present is represented as where the indeterminate nature of potential physical events changes to a definite outcome, and even the nature of the future spacetime is uncertain until it is determined at that time, along with the physical events that occur in it. However, the transition from present to past does not take place on specific spacelike surfaces, being determined by any universal time defined by such spacelike surfaces; rather *it takes place pointwise at each spacetime event*. The past is determined at each event, the future is undetermined, and the [here]-now is a moment of passage from the one state to the other in a point wise way. It is however convenient to consider the *evolution as taking place along timelike or null world lines* related to ongoing physical processes. Indeed this is strongly suggested both by the way that time is determined in General Relativity as a path integral along timelike world lines, and also by the way that physical effects that determine what happens take place along histories of matter, represented by timelike worldlines or null worldlines. Which is more important will depend on the epoch in the history of the universe; at all recent times it is timelike worldlines that count rather than null (when null based processes are important, usually many null world lines will be involved and the effect averages out to a timelike based process).

But then the question is which world line should be chosen? The potential problem is the arbitrariness in the choice of the world lines. There seem to be two choices: either
 - *we regard the evolution of time as being allowed to take place along any world lines whatever, none being preferred*, or
 - *the evolution of time takes place along preferred world lines, associated with a symmetry breaking that leads to the emergence of time*, and presumably to the arrow of time [11,12].

In the latter case, the question is, which world lines are the key ones that play this crucial physical role? In specific realistic physical situations, there will be preferred world lines associated with the average motion of matter present, as in the case of cosmology, and there will be preferred time surfaces associated with them if the matter flow is irrotational. This occurs in particular in the case of the idealised Robertson-Walker models of standard cosmology, where as long as matter is present, there exist uniquely preferred irrotational and shear-free worldlines that are eigenvectors of the Ricci tensor. These then form a plausible best basis for description of physical events and the evolution of matter: there is a unique physical evolution determined along each such a family of world lines with its associated unique time surfaces, which are invariant under the space time symmetries. Then one might propose that the evolution of time is associated with these preferred timelike world lines and perhaps associated spacelike surfaces, being an emergent property associated with the broken symmetries represented by these geometrical features in curved spacetimes.

But there may be several competing such choices of world lines in more realistic cases, for example realistic perturbed cosmological models such as are needed for structure formation studies will have multiple matter components present with differing 4-velocities. For this reason, we might rather consider the evolution as taking place along *arbitrary families of world lines*, corresponding to the freedom of choice of the shift vector in the ADM formalism for General Relativity [13]. A key result then is that no unique choice for these world lines needs to be made in the standard General Relativity situation with simple equations of state; the ADM theory says *we locally get same result for the evolving spacetime, whatever world lines are chosen.* You can choose any time lines you like to show how things will have evolved at different places (that is, on different observer's world lines) at different times (that is, at various times along those world lines). But this view has no foundationally preferred status: you could have chosen different world lines, corresponding to different shift vectors, and a different relation between times on the world lines, corresponding to different choices of the lapse function; the resulting four dimensional spacetime is the same. In any specific situation, some of those descriptions will be more natural and easier to use and understand than others; but this is just a convenience, and any other surfaces and world lines could have been chosen.

Thus in the classical GR case, we get a consistent picture: things are indeed as we experience them. Time rolls on along each world line; the past events on a world line are fixed and the future events on each world line are unknown. Spacetime grows: even the spacetime structure itself is to be determined as the evolution takes place (Figure 2). The metric tensor determines the rate of change of time with respect to the coordinates, for this is the fundamental meaning of the metric [14]. A gauge condition determines how the coordinates are extended to the future. Conservation equations plus equations of state and associated evolution equations determine how matter and fields change to the future, including the behaviour of ideal clocks, which measure the passage of time. The field equations determine how the metric evolves with time, and hence determine the future space-time curvature. It is convenient to introduce local coordinates in order to determine how this works, involving a splitting of spacetime into space and time as in the ADM formalism [13], with evolution along the coordinate lines introduced as determined by the shift function. The whole fits together in a consistent way, determining the evolution of both space-time and the matter and fields in it, as is demonstrated for simple equations of state by the existence and uniqueness theorems of general relativity theory ([14]:226-255).

**3.2 Semi-classical gravity**
Thus no unique choice needs to be made for the conventional ADM formalism, which is deterministic; for these standard theorems assume classical deterministic physics rules the micro-world. But the whole point of this paper is that most models are *not* deterministic, irreversible unpredictable processes and emergent properties will take part in determining space time curvature. A realistic extension of the above account to the semi-classical case will take into account quantum uncertainty in the evolution of the matter and fields, giving a probability for the future evolution of particles, fields, and hence for spacetime, rather than a definite prediction. Quantum evolution will determine the actual outcome that occurs in a probabilistic way. But then the problem is, if two choices of world lines are made *in two different indeterministic futures*, then it is probable that the two evolutions will not agree. In such a scenario we would have something more like Wheeler's "many fingered time" – the different proper times along arbitrary world lines do not knit together to form a global concept of time that is meaningful in determining a unique evolution along all world lines. How then can an evolving block universe emerge from this situation?

The fundamental observational fact is,

- *In the real universe domain we actually inhabit, a unique classical space-time structure does indeed emerge at macro scales from the underlying physics*

Hence whatever integrability conditions are needed for this to happen, they do indeed occur in our universe domain, and we are able to describe what happens via an evolving block universe picture. It is for this reason that we are able to regard special and general relativity as successful theories in their appropriate contexts within the local universe domain in which we live. This is related to the emergence of the classical from the underlying quantum theory [5,15].

**5: Implications**

In some circumstances, classical physics emerges from quantum theory as an approximation in which physical processes are time reversible. At the macro level the outcome will almost always be irreversible, both because of dissipative processes such as friction that emerge at higher levels, and because of the emergence of complex systems where the flow of entropy is an inexorable feature of life, indeed one of the most fundamental features of chemistry and biology (as represented by the second law of thermodynamics). The exceptions are some cases in astronomy where friction is negligible, such as planetary motion. But even then this reversibility only holds for a limited period of time; the processes of planetary formation were strictly irreversible.

Thus what is needed is *a fundamental change of view: the default state in physics is not a time reversible flow of events with no distinguished present; it is an ongoing time irreversible flow with dramatically distinguished past, present, and future*. There are some conditions where this is approximated well by a time reversible flow, where the present is not particularly different from the past and the future, but this is not the fundamental underlying situation, it is an emergent approximation that is only sometimes valid.

Physics should be framed with this standpoint at its foundations, rather than being based on the view that fundamental physics is time reversible. The basic issue then is not why macroscopic situations involving chemistry and biology are time irreversible, it is why in some cases time reversible physics is a good approximation. At the micro scale, under some special circumstances even quantum measurements are reversible [16]; but this is only in very exceptional situations. At the macro scale, coarse graining ensures that what actually occurs is irreversible [2]; in complex systems, the physical structure that determines what happens embodies a specific coarse-graining that ensures the flow of time is a fundamental feature of life. Furthermore, one should note that even when micro processes that take place are themselves time reversible, the actual event that happens occurs in one direction of time: electrons collide with positrons and emit photons in a specific case, for example, rather than the other way round. What actually occurs rolls on in a uni-directional way, even when the physics is time reversible. The time symmetry of the equations is broken in physical reality.

This understanding undermines approaches to determining the arrow of time based on comparing the far future with the far past [11], as in the Wheeler-Feynman [17] and Penrose [2] approaches; it means such ways of trying to determine the arrow of time are simply not applicable. You can't integrate over the far future if that future does not yet exist and its nature is not yet determined. Nor can one (in this context) sensibly propose situations where the future retrospectively determines the present or past (e.g. [18]: "We explore the idea that the dynamics of the inflationary multiverse is encoded in its future boundary, where it is described by a lower dimensional theory"). Instead the arrow of time is determined here and now through decoherence and the time irreversible nature of the quantum measurement process, which are not captured by the Schrödinger, Dirac, or Wheeler-de Witt equations (and so are ignored in Barbour's analysis [1]), and inevitably lead to a progress of time in biology and in our minds. The direction of time in which growth of the EBU inexorably takes place is locally determined via decoherence and effective collapse of the wave function, leading to classical behaviour [3,5,15] with a well-determined arrow of time.

**6: Conclusion**

The view proposed here is that spacetime is extending to the future as events develop along each world line in a way determined by the complex of causal interactions; these shape the future, including the very structure of spacetime itself, in a locally determined (pointwise) way. Spacetime is an Evolving Block Universe that continues evolving along every world line until it reaches its final state as an unchanging Final Block Universe. One might say that then time has changed into eternity. The future is uncertain and indeterminate until local determinations of what occurs have taken place at the space-time event `here and now', designating the present on a world line at a specific instant; thereafter this event is in the past, having become fixed and immutable, with a new event on the world line designating the present. There is no unique way to say how this happens relatively for different observers; analysis of the evolution is conveniently based on preferred (matter related) world lines rather than time surfaces. However in order to describe it overall, it will be convenient to choose specific time surfaces for the analysis, but these are a choice of convenience rather than necessity.

This may be difficult to implement in physical theory, but it is actually the way things work; present theoretical physics understanding simply does not adequately represent it. It is probably profoundly implicated in the issues of quantum measurement and decoherence, which are time-irreversible processes that may underlie the way that time moves on. The need is to revisit these processes in a way that takes the Evolving Block Universe seriously. And fundamentally, this process view of time should be built into the foundations of any quantum gravity theory, properly related to the quantum measurement issue and the continual occurrence of decoherence. That this emergent block universe viewpoint is possible in quantum gravity theories is shown for example by spin foam proposals [20].

Thus a key element is how this proposed viewpoint relates to quantum gravity. But equally important is how it relates to the emergence of complexity. On the one hand, this is only possible because the initial state of our expanding universe domain was a low entropy state [2]; on the other, it plays itself out in the interplay between bottom-up and top-down interactions that enables true complexity, such as life, to emerge [21] and experience the overwhelming dominance of the flow of time in daily life. The way that this happens remains a fruitful issue of exploration. But one thing is clear: both the entire Darwinian process of evolution through which we come into existence, and the processes by which we read this article, depend on the flux of time. You would not exist and have the ability to read this article if the view proposed here (and expounded in more detail in Ref. [6]) was not a correct description of the way things are.

Cape Town
2008-12-01

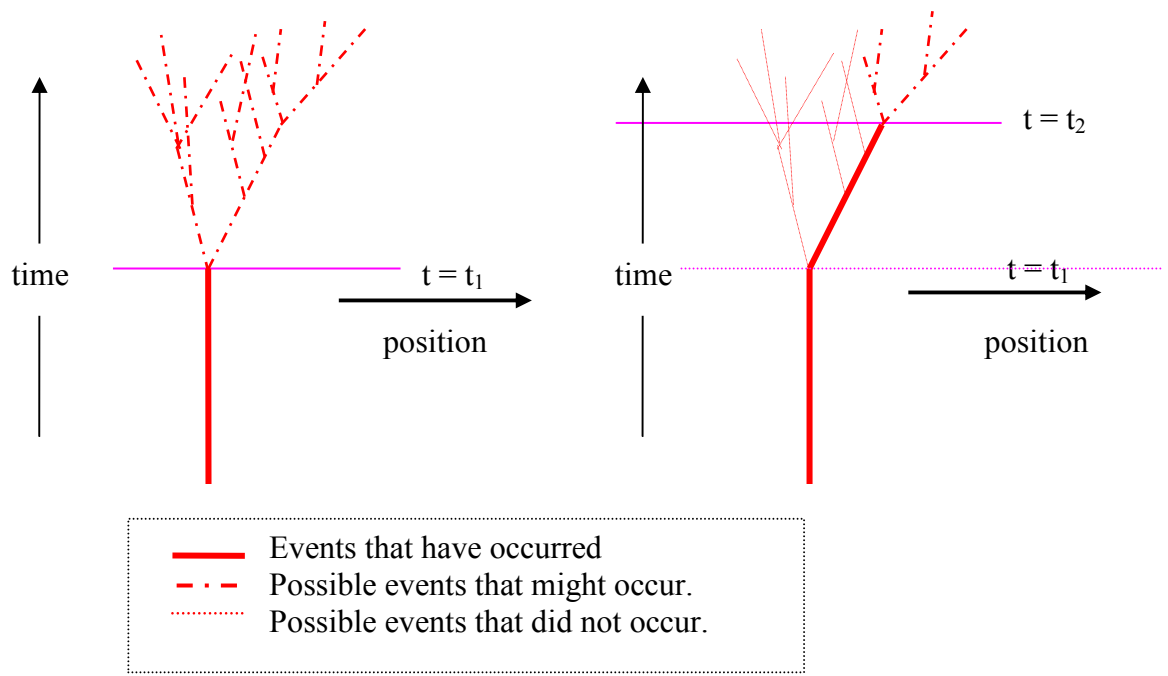

**Figure 1**: *Motion of a particle world line controlled in a random way, so that what happens is determined as it happens. On the left events are determined till time $t_1$ but not thereafter; on the right, events are determined till time $t_2 > t_1$, but not thereafter.*

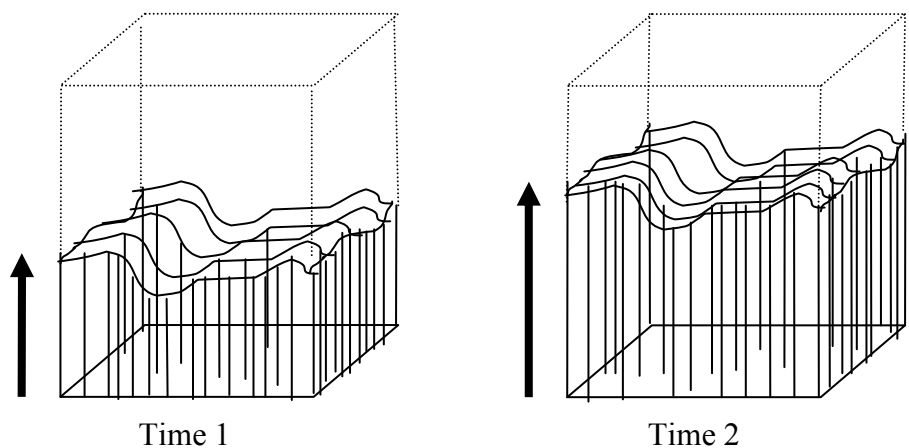

Time 1    Time 2

**Figure 2:** *An evolving curved space-time picture that takes macro-phenomena seriously. Time evolves along each world line, extending the determinate spacetime as it does so (what might be changes into what has happened; indeterminate becomes determinate). The particular surfaces have no fundamental meaning and are just there for convenience (we need coordinates to describe what is happening). You cannot locally predict uniquely to either the future or the past from data on any `time' surface (even though the past is already determined). This is true both for physics, and (consequently) for the spacetime itself: the developing nature of spacetime is determined by the evolution (to the future) of the matter in it.*